\documentstyle [pre,aps,epsfig,multicol]{revtex}
\newenvironment{mytitle}{\begin{center} \large \bf }{\\ [.1in]\end{center}}
\newenvironment{myauthor}{\begin{center} \large }{\\ [.1in]\end{center}} 
\newenvironment{myinstit}{\begin{center} \large \it}{\end{center}}
\begin{document}

\thispagestyle{empty}

\begin{mytitle}
Chaotic Electron Motion in Superlattices. Quantum-Classical Correspondence of 
the Structure of Eigenstates and LDOS.
\end{mytitle}

\begin{myauthor}
G. A. Luna-Acosta,  J. A. M\'endez-Berm\'udez, and F. M. Izrailev
\end{myauthor}

\begin{myinstit}
Instituto de F\'{\i}sica, Universidad Aut\'onoma de Puebla, Puebla, Pue. 
72570, Apartado Postal J - 48, M\'exico.
\end{myinstit}

{\large 
\begin{center}{\bf Abstract} 
\end{center}

We investigate the classical-quantum correspondence for particle motion in a superlattice
in the form of a 2D channel with periodic modulated boundaries. 
Its classical dynamics undergoes the generic transition to chaos of 
Hamiltonian systems as the amplitude of the modulation is increased. We show 
that for strong chaotic motion, the classical counterpart of the 
structure of eigenstates (SES) in energy space reveals an excellent 
agreement with the quantum one. This correspondence allows us to 
understand important features of the SES in terms of classical 
trajectories. We also show that for typical 2D modulated waveguides there exist, at any
energy range, extremely localized eigenstates (in energy) which are practically unperturbed by the
modulation. These states contribute to the strong fluctuations around the classical SES. The 
approach to the classical limit is discussed.


\begin{multicols}{2}

A variety of important physical systems are periodic, either in space;  such as crystals
 and superlattices; in time, like periodically  forced mechanical devices; or in both, as
 in particle motion in a superlattice subject to an ac electric field. The analysis of such
 systems has lead to the understading of well known fundamental phenomena, e.g., energy band structure
 of solids and resonances. On the other hand, during the last decade much attention have been paid to
to study the classical-quantum correspondence (CQC) of classically chaotic systems.
  In [1] we give references to some interesting physical time-periodic chaotic systems.
 One of the challenges is to identify the ways in which classical chaos may be manifested
in the quantum regime [2]. There are well established tools to characterize a quantum
 classically chaotic system. The most commonly used relies on the analysis of their spectra
 (e.g., level spacing distribution) in the context of the random matrix theory 
conjecture [3].
See also [4] for other signatures of "quantum chaos".

Here we explore the CQC of the particle motion in
a superlattice in the form of a 2D electron waveguide with periodically modulated
boundaries. Two possible experimental realizations are: 1) a film whose thickness is a periodic
function of one of the coordinates and 2) a periodically modulated mesoscopic
electron channel in the
ballistic regime (see [5,6] and references therein). We analyze the problem
by studying the structure of the eigenfunctions (SEF) and local
 density of states (LDOS) [7]. 
This novel approach has been applied recently to a variety of systems [8]. 
The basic idea is as follows. Consider a non-integrable Hamiltonian $H=H_0 + V$, where $V$is the
non-integral perturbation to the integrable Hamiltonian $H_0$. Let the eigenstates of
$H$ and $H_0$ be denoted, respectively, by $\Psi^{\alpha}$ and $\phi_j$.  
We can then form the matrix $w_l^\alpha \equiv |C^{\alpha}_l(k)|^2$, where $C^{\alpha}_{\ j}
 = < \Psi^{\alpha}\mid \phi_j>$. The rows (columns) 
of $w_l^\alpha$ show how a
 specific perturbed eigenstate $|\alpha>$ (unperturbed eigenstate $\phi>$) is expanded in the
unperturbed basis $|l>$ (perturbed basis $\alpha>$). For our purposes it is essential, as will become
clear below, to energy-order the unperturbed as well as the perturbed states; that
is, $E^{\alpha+1}\geq E^{\alpha}$ and $E^0_{l+1}\geq E^0_l$, where
 $E^{\alpha}$ and 
$E^0_l$  are the energy spectra of the perturbed and unperturbed systems,
 respectively; $\alpha,l=1,2...$.
The structure of eigenstastes SES is defined as [7],
\begin{equation}
W(E_l^0 \mid E^\alpha)=
\sum_{\alpha'} \bar{w}_l^{\alpha'} \,\delta(E_l^0-E^{\alpha'}),
\label{sef}
\end{equation}
where $\bar{w}_l^{\alpha'}=w_l^{\alpha'}/N$ with $N$ as the number
of eigenstates $|\alpha'>$ in the vicinity of a given $\alpha$. 
Eq. (1), seen as a function of the unperturbed energy $E_l^0$
gives the structure of eigenstates with total energy close to $E^\alpha$. 
The local density of states LDOS, also known as the strength function, defined as
$\omega(E^{\alpha} \mid E^0_l) = \sum_{l'} \bar{w}_{l'}^\alpha \,
\delta (E^{\alpha}-E^0_{l'})$, is also used to study the quantum-classical correspondence
[6,8,9]. Here, for lack of space we shall not discuss it.

Now, the reason these two quantities, SES and LDOS, help us tackle the quantum-classical
correspondence is that they have well defined classical counterparts [7].
 Since
$C^{\alpha}_l =<\Psi^{\alpha}\mid\phi_l>$ as a function of $l$ is the
{\it projection} of the perturbed state onto the states of the
unperturbed system, the classical counterpart of $w_l^\alpha =
\mid C^{\alpha}_l\mid^2$ as a function of energy $E_l$ can be
defined  as the {\it projection} of the total Hamiltonian $H$ onto
the unperturbed one $H^0$, where $H=H^0 +V$ with $V$ the
perturbation. This can be numerically
done by substituting the trajectories $\Phi(t)\equiv (x(t), y(t),
p_x(t), p_y(t))$ generated by $H$ with energy $E$
into $H^0$. Since the unperturbed
energy $E^0(t)$ along these trajectories varies in time, it fills
the so-called {\it energy shell} characterized by its width
$\Delta E$. For chaotic total Hamiltonians $H$, the
classical analog $W_{cl}(E^0 \mid E)$ of the quantum SES is the probability distribution
constructed from $E^0(t)$. See below for details. 

 Similarly, the classical analog of LDOS
can be obtained by performing the inverse operation, namely, projecting the dynamics
of $H^0$ onto the Hamiltonian function $H$. 

This approach assumes that the Hamiltonian could be separated into
unperturbed and perturbed parts. For billiards, like our system, the non-integrability comes
from the boundary conditions; the Hamiltonian operator, being simply the kinetic energy, is the same
for both perturbed (periodically modulated) and unperturbed (flat) channel. To overcome this problem,
a transformation to curvilinear coordinates can be performed such that both, perturbed and unperturbed
systems have the same boundary conditions. Consequently the effects of the boundary appear in the
new coordinates as an interaction potential between the two degrees of freedom (see details in
[9]).

Since the Hamiltonian is periodic, the energy eigenstates satisfy Bloch's theorem.
Although most of the results will be independent of the details of the profile, we need to specify
a particular one. As in [5,6,9] we 
define the top boundary by $y=d+a \cos(x)$ and the bottom boundary by $y=0$. 
The first classical dynamical studies of this system appear in [10,11]. The finite length version
this system was analyzed in [12] as a model of a mesoscopic electron
waveguide, where a transport signature of chaos in the behavior
of resistivity was established. Ketzmerick [13] also used it to study certain
{\it quantum} transport properties through ballistic cavities. 
Moreover, the analysis of the band-energy spectrum for an {\it infinitely} long
rippled channel [5] in the cases of mixed and global
classical chaos, yields insight into the universal
features of electronic band structures of real crystals
[14]. See also [15]. Results obtained for 
this particular system are applicable to a large class of 
systems, namely non-degenerate, non-integrable Hamiltonians.
We limit this report to the case of global classical chaos which occurs in wide 
($d>2\pi$) channels [5]. The geometrical parameters used here are $(a/2\pi,d/2\pi)=(0.12,0.5)$.
Fig. 1 shows the unperturbed energy $E^0(t)$ obtained as explained above.
The distribution of energies on the right of Fig. 1 is, by definition, the classical 
counterpart of SES, obtained by averaging $E^0(t)$ over time.
\begin{figure}[htb]
\begin{center}
\epsfig{file=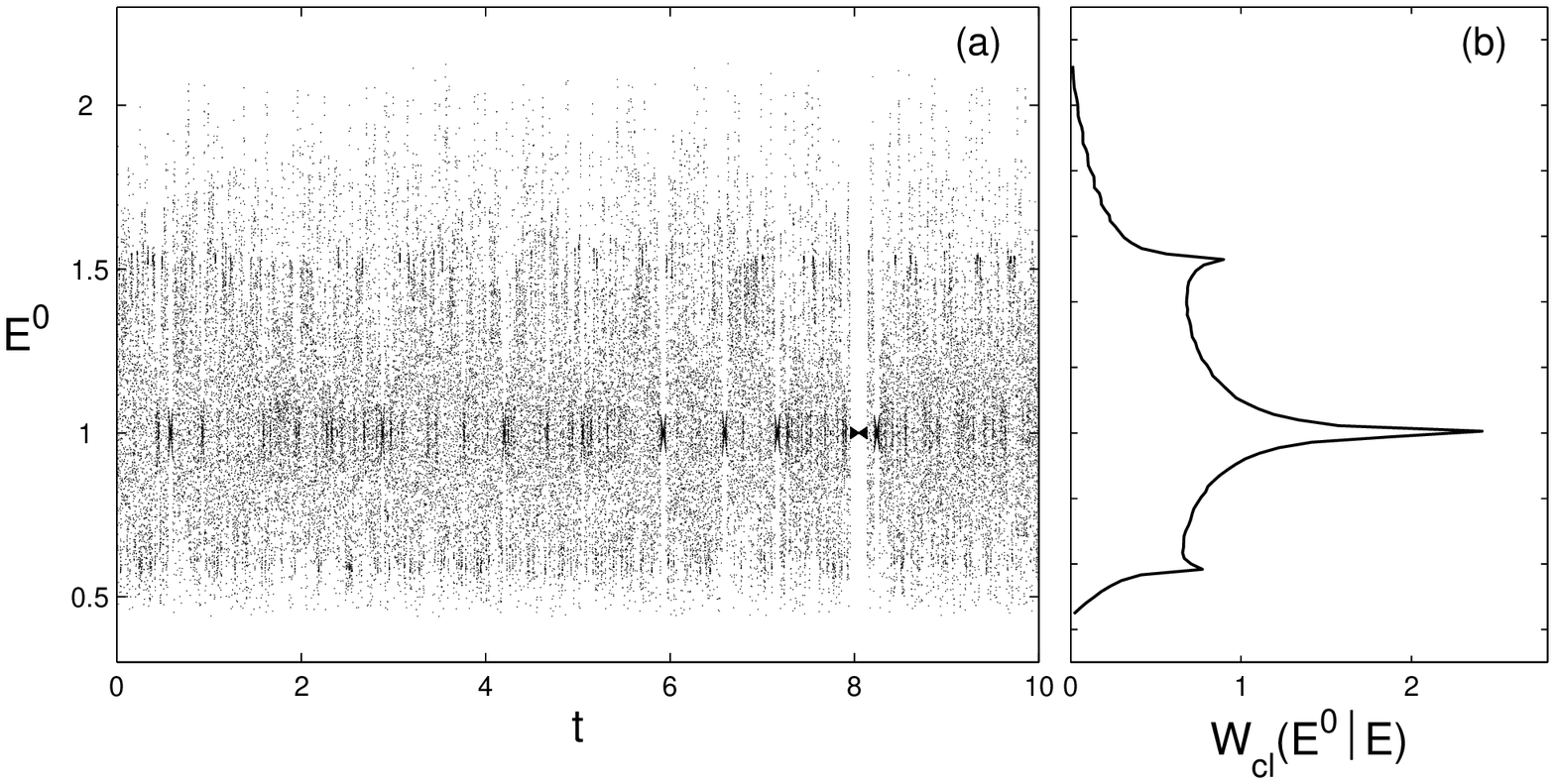,width=3in,height=1.3in}
\label{weaal0}
\end{center}
\end{figure}
\begin{center}
\vspace{-0.28in}
{\normalsize FIG. 1 a) $E^0(t)$ as a
function of time (in arbitrary units) for $E=1$. b) Classical SES
$W_{cl}(E^0 \mid E)$.}
\end{center}
\begin{figure}[htb]
\begin{center}
\epsfig{file=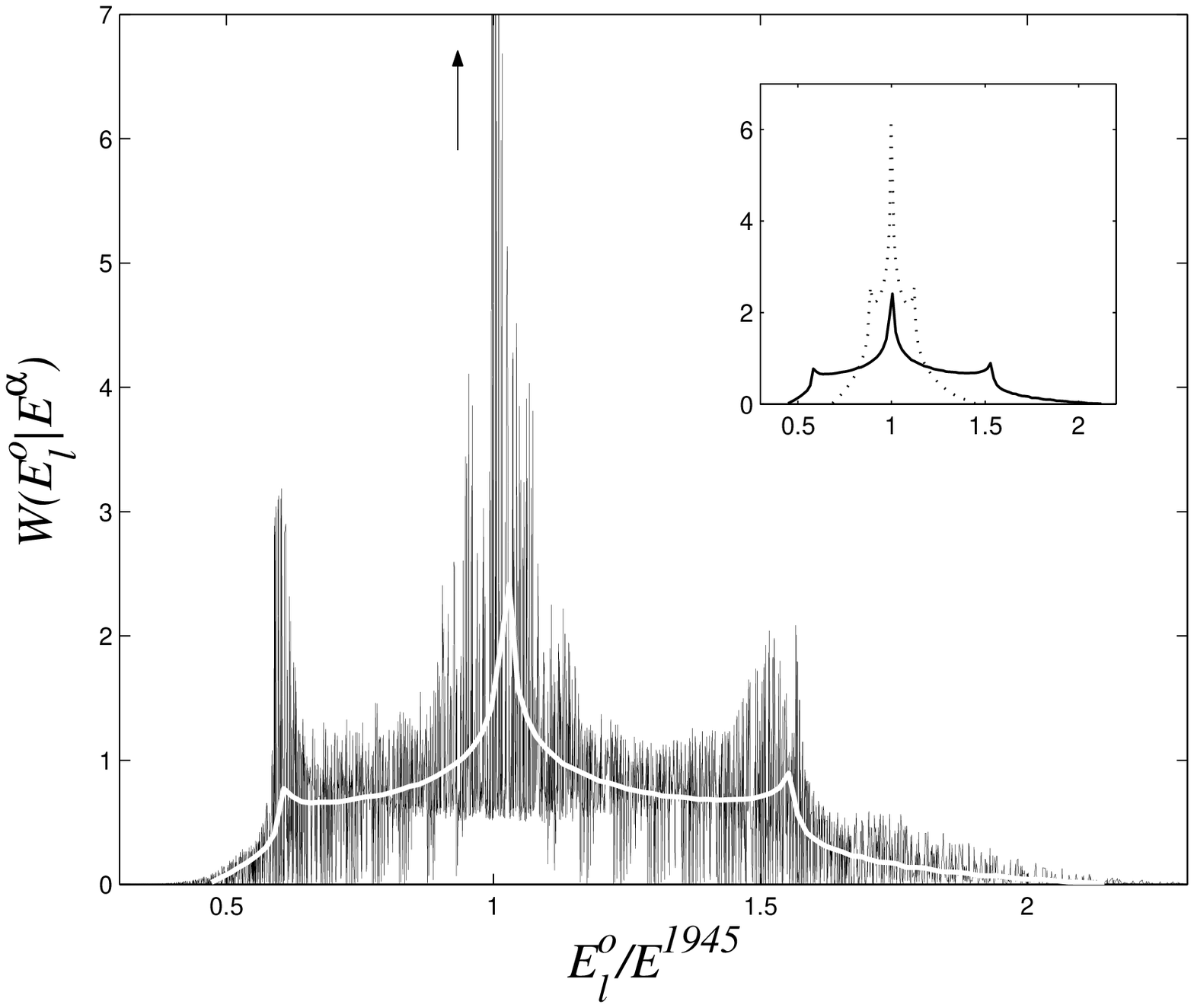,width=3in,height=2.4in}
\label{weaal0}
\end{center}
\end{figure}
\begin{center}
\vspace{-0.28in}
{\normalsize FIG. 2 Structure of eigenstates. Quantum SES (black line); Classical SES (white
line). The average for the SES is over $1900<\alpha<1990$. Inset: classical SES for ($a/2\pi, d/2\pi) 
= (0.06,1)$ (dotted) and ($a/2\pi, d/2\pi) = (0.12,0.5)$ (continuous).}
\end{center}
Fig. 2 compares the quantum and classical SES. We see that the
classical SES agrees well with the average shape of quantum SES. {\it I.e.}, the classical SES 
predicts: 1)
the appearance of three prominent peaks at specified values, 2) the width of the distribution 
(the energy shell); and 3) its assymmetry about 1. The inset compares classical SES for two 
channels defined by different geometrical parameters but both displaying global chaos in phase
space. Thus, the classical and quantum SES can detect {\it dynamical differences not revealed
by the standard tools}, e.g., Poincar\`{e} maps, and thus serves to complement the 
characterization of dynamical
systems.
In fact, detailed analysis [9] demonstrates that the left (right) side peak is formed
by trajectories dwelling near the unstable (stable) fixed period-one fixed point, whereas the central
peak is due to grazing trajectories.

The strong fluctuations of the quantum SES are due to: 1) the fact that the system is still not deep
in the semiclassical regime (although the level numbers are high) and 2) the existence of extremely
localized and sparse (in energy space) states. See below and [9] for details. 

\begin{figure}[htb]
\begin{center}
\epsfig{file=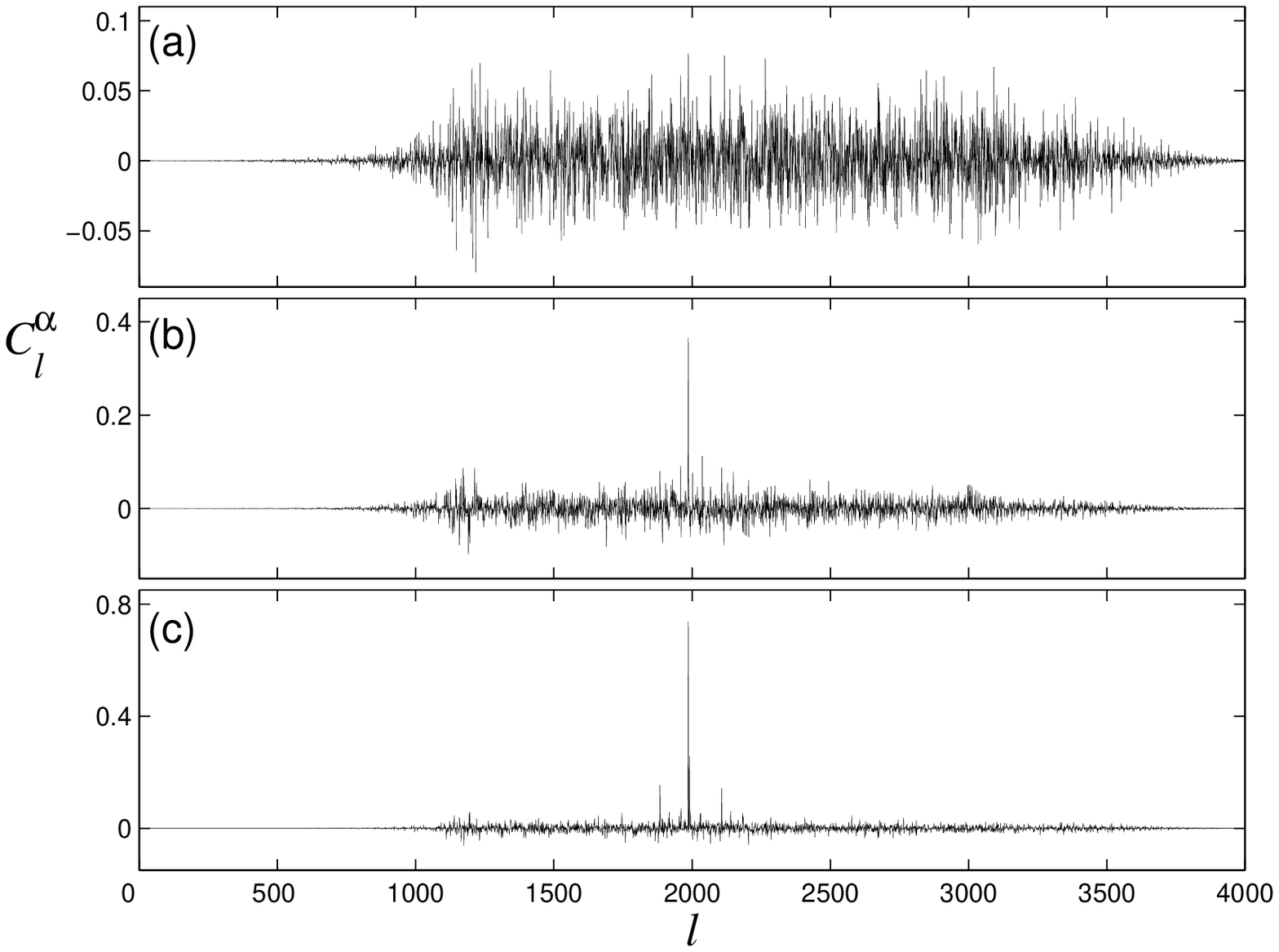,width=3in,height=2.5in}
\label{weaal0}
\end{center}
\end{figure}
\begin{center}
\vspace{-0.28in}
{\normalsize FIG. 3 Typical eigenfunctions. a) $\alpha$=1984,
b) $\alpha$=1985, and c) $\alpha$=1986.}
\end{center}

Fig. 3 shows 3 consecutive eigenstates
in the energy-ordered unperturbed basis, illustrating the kind of eigenstates that
 can occur typically: extended, sparse, and localized in the energy representation. A quantitative
characterization is given by various localization measures, such as the entropy localization length
$l_H$, the inverse participation ratio $\l_{ipr}$, and the mean square root $\l_\sigma$. See [6] for their definitions. In Fig. 4 we present $l_H$ as a function of the
perturbed state. It shows wild fluctuations, due to the existence of the localized and sparse states
mentioned above. The extremely localized states are those that are practically the same as the
unperturbed (flat channel) states with transversal mode number $m=1$. It can be shown 
that the number of these states $N_{m=1}$, relative to
the total number of states $N(E)$, up to energy $E$, is given by $N_{m=1}/N(E)=
4/\pi \sqrt {E}$.

\begin{figure}[htb]
\begin{center}
\epsfig{file=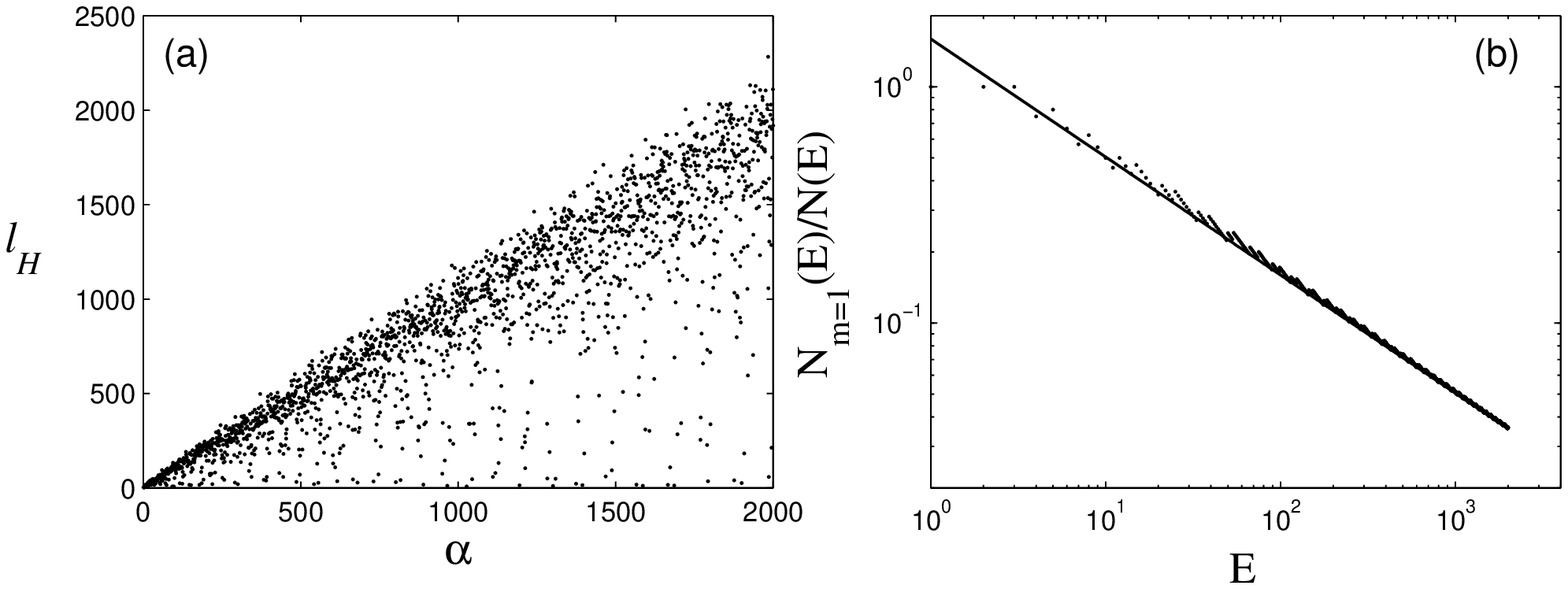,width=3in,height=1.5in}
\label{weaal0}
\end{center}
\end{figure}
\begin{center}
\vspace{-0.28in}
{\normalsize FIG. 4 (a) Entropy localization
length $l_H$ as a function of $\alpha$. (b)  
$N_{m=1}(E)/N(E)$ as a function of $E$ (dots) for the first $2000$ states. 
The solid line is the analytical estimate $4/\pi \sqrt {E}$.} 
\end{center}
The fact that $N_{m=1}/N(E)$ decays as $1/\sqrt{E}$ implies that 
in the strict classical limit the fluctuations, observed in Fig. 2
for the quantum SES, will vanish, however, the approach to this limit is surprisingly slow.
 Furthermore, it can be shown that $N_{m=1}=2\pi \sqrt{E}/d$, where $d$ is
the width of the channel, in units of the period of the ripple. This indicates that
 localized states continue to appear at all energies, even in the classical limit!. That is, 
one can always find states that remain practically unperturbed by the modulation. The importance
of these extremely localized states in finite electron waveguides with modulated or corrugated
boundaries is addressed in [16].\\

{\bf Acknowledgements}. We acknowledge support by  CONACyT grant, No. 26163-E, and 
NSF-CONACyT, grant No. E120-3341.\\

{\normalsize 
\noindent [1] O.Agam, S. Fishman, and R.E. Prange, Phys. Rev. A {\bf 45},6773(1992);S. Ree
and L. E. Reichl, Phys. Rev. {\bf E53},1228 (1996); M. Latka, P. Grigolini, and B. West, 
Phys. Lett. {\bf A 189},145 (1994); M. Henseler, T. Dittrich, and K. Richter, Europhys. Lett.
{\bf 49},289 (2000).

\noindent [2] F.M. Izrailev, Phys. Rep.{\bf 196}, 299 (1990); F. Haake, {\it Quantum signatures of Chaos} 
(Springer-Verlag, Berlin, 1991); L.E. Reichl, {\it The Transition to Chaos in Conservative Classical Systems: Quantum Manifestations}
 (Springer-Verlag, new York, 1992); H-J. Stockmann, {\it Quantum Chaos: An Introduction} (Cambridge Univ. Press, 1999, UK).

\noindent [3] G. Casati, F. Valz-Gris, and I. Guarneri, Lett. Nuovo Cimento. {\bf 28}(1980)279;
O.Bohigas, M.-J. Giannoni, and C. Schmidt, Phys. Rev. Let.{\bf 52}, 1(1984).

\noindent [4] F. Leyvraz, J. Quezada, T.H. Seligman, and M. Lombardi, Phys. Rev. Lett.{\bf
67},2921 (1991); W.H. Zurek and J.P. Paz, Physica D {\bf 83},300 (1995);and T. Dittrich, B. Mehlig, H. Schanz, and U. Smilansky, Phys. Rev. E {\bf 57}, 359 (1998).

\noindent [5] G.A. Luna-Acosta, K. Na, L.E. Reichl, and A. Krokhin, Phys. Rev. E{\bf 53}, 3271 (1996).

\noindent [6] G.A. Luna-Acosta, J.A. M\'endez-Berm\'udez, and F.M. Izrailev, Phys. Lett. {\bf A 274},
192 (2000).

\noindent [7] G. Casati, B, Chirikov, I. Guarneri, and F.M. Izrailev, Phys. Lett. {\bf A 223} (1996)430.

\noindent [8] F. Borgonovi, I. Guarneri, and F.M. Izrailev, Phys. Rev. {\bf E 57}(1998) 5291; 
Weng-ge. Wang, F.M. Izrailev, and G. Casati, Phys, Rev, {\bf E 57}(1998) 323;
L. Benet, F.M. Izrailev, T. H. Seligman, A. Su\'arez-Moreno,Phys. Lett {\bf A 277},87
(2000); and L. Meza-Montes, F.M. Izrailev, and S. E. Ulloa, Phys. Stat. Sol {\bf b 271},721(2000).

\noindent [9] G. A. Luna-Acosta, J. A. M\'endez-Berm\'udez, and F. M. Izrailev, Phys. Rev. E {\bf 64} 036206 (2001).

\noindent [10] J. L. Tennyson, p. 158 in {\it Nonlinear Dynamics and
the Beam-Beam Interaction}. M. Month, and J. C. Herrera (eds.), AIP
Conference Proceedings No. 57. American Institute of Physics, New York, 
1979.

\noindent [11]  A. J. Lichtenberg, and M. A. Lieberman, {\it Regular and
Chaotic Dynamics}, 2nd ed., (Springer-Verlag, New York, 1992),
Sec. (6.1b).

\noindent [12] G. A. Luna-Acosta, A. A. Krokhin, M. A. Rodriguez, and
P. H. Hernandez-Tejeda, Phys. Rev. B {\bf 54},11410 (1996).

\noindent [13] B. Huckestein, R. Ketzmerick, and C. H. Lewenkopf,
Phys. Rev. Lett., {\bf 84}, 5504 (2000).

\noindent [14] E. R. Mucciolo, R. B. Capaz, B. L. Altshuler, and
J. D. Joannopoulos, Phys. Rev. B {\bf 50}, 8245 (1994).

\noindent [15] T. Dittrich, B. Mehlig, H. Schanz,
and U. Smilansky, Phys. Rev. E {\bf 57}, 359 (1998); 
V. Ya. Demikhovskii, S. Yu. Potapenko, and A. M. Satanin,
Fiz. Tekh. Poluprovodn. {\bf 17}, 213 (1983) [Sov. Semicond. {\bf
17}, 137 (1983)]; C. S. Lent and M. Leng,
J. Appl. Phys. {\bf 70}, 3157 (1991).

\noindent [16] G. A. Luna-Acosta, J. A. M\'endez-Berm\'udez, P. \v{S}eba, and K. N. Pichugin, Phys. Rev. E, in press (2001); idem, cond/mat-0107418.
}

\end{multicols}

}

\end{document}